# Generalized Green-Ampt Approach to 1D Oscillatory Flows in Partially Saturated/Unsaturated Media: Capillary Effects in Beach Hydrodynamics (Semi-analytical & Numerical studies)

**Khalil Alastal[1*], Rachid Ababou[1], Dominique Astruc[1]**

[1]Institut de Mécanique des Fluides de Toulouse,
2 Allée du Professeur Camille Soula, F-31400 Toulouse, France.
Email: kalastal@imft.fr, ababou@imft.fr, astruc@imft.fr

**\*Corresponding Author**
Khalil ALASTAL: kh.astal@gmail.com ; khalil.alastal@imft.fr ;
kastal@iugaza.edu.ps

Tel: +33(0)5.34.32.28.25
Institut de Mécanique des Fluides de Toulouse (IMFT),
2 Allée du Professeur Camille Soula, 31400 Toulouse, France

# Generalized Green-Ampt approach to 1D oscillatory flows in partially saturated/unsaturated media: capillary effects in beach hydrodynamics (semi-analytical and numerical studies)

Khalil Alastal, Rachid Ababou, Dominique Astruc

**Abstract**

Semi-analytical multi-front solutions of water table response due to periodic forcing in a partially saturated vertical porous column are developed, tested and compared to finite volume solutions of the Richards equation for partially saturated / unsaturated flow with non linear water retention and conductivity curves ($\theta(h)$, $K(h)$). The multi-front solutions are useful for capturing parametrically the frequency response of the vertical column to tidal oscillations while taking into account both capillary and gravitational effects. Vertical oscillations are examined, accounting for unsaturated flow above the oscillating water table as well as saturated flow below it. The multi-front models are conceived as successive generalizations of the Green-Ampt piston flow approach. The single front model is an "inverted" Green-Ampt model, with an abrupt front separating the saturated and dry regions. It is adapted to the case of an oscillatory pressure imposed at the bottom of the column (rather than a fixed pressure imposed at the top). The N-front models ($N \geq 2$) further generalize this concept, using a ($\theta(h)$, $K(h)$) parametrization to take into account the capillary properties of the unsaturated medium. The resulting systems of ODE's are non linear with time variable coefficients (non autonomous ODE systems). The solutions obtained for $N \approx 10$ fronts are satisfactory both in terms of water table fluctuations and moisture profiles, even for fine grained soils (Guelph Loam). They are computed much faster than space-time discretized solutions of the non linear Richards PDE. For sandy soils, even the 2-front solution (N=2) is satisfactory in terms of water table response $Z_s(t)$. The 2-front model itself is a significant improvement on the single front Green-Ampt model, and it appears potentially useful for analyzing the response of unsaturated flow systems under various types of oscillatory and transient forcing. Overall, the N-front method is useful for exploring the frequency response of the water table to tidal forcing. To illustrate this capability, we discuss the results of a parametric study of mean water table height *vs.* frequency for the Guelph Loam.

## 1. Introduction and literature review

In coastal hydrodynamics, oscillations of groundwater flow and water table elevation in beaches have been recognized to influence the morphological processes and sediment transport of the swash zone (erosion and accretion) [1,2]. The oscillatory flow regime can also control the biological conditions, nutrient cycling and contaminant movement near the water table [3,4]. These oscillations also affect the stability of structures founded on soils and sands [5]. Such oscillation phenomena also take place in river banks near estuaries, in harbor dykes, etc. The porous media involved in these oscillatory phenomena may have various grain size distributions. In many cases, due to fine grained features, capillary effects intervene significantly, and one must account for the time variable unsaturated wetting and drainage phenomena above the oscillating water table.

Thus, a reliable and simple model that predicts the space-time response of groundwater flow and water table elevation in the presence of unsaturated capillary effects, would be very useful for studying the nonlinear response of groundwater flow and water table elevation to, say, periodic forcing. In this paper we will focus specifically on the case of tidal forcing at the bottom of a vertical column, without taking into account wave hydrodynamics, but taking into account both saturated and unsaturated flow dynamics in the porous column.

Many models have been developed to predict the water table fluctuations in response to periodic forcing [5,6,7,8,9,10,11]. Most of the proposed models are based on the Dupuit-Boussinesq equation of plane groundwater flow [12,13], which is derived from Darcy's law and mass conservation assuming (i) vertically hydrostatic pressure (plane flow), and (ii) instantaneous wetting/drainage of the moving water table (the unsaturated zone is totally dry at all times). The models based on (i) and (ii) can predict space-time propagation of water table fluctuations horizontally (Zs(x,y,t)), but they are limited to planar flow within the saturated region, and also, they totally neglect any capillary effect on water table fluctuations.[1]

Recent laboratory and field studies show that capillarity affects water table dynamics over a wide range of frequencies, including tidal frequencies [14,15,16,17]. As the water table fluctuates, the pressure distribution above the water table will change, and thus an apparent local water exchange across the water table occurs [18].

Parlange and Brutsaert [19] proposed a correction of the Dupuit-Boussinesq equation to take into account capillary effects above the water table. Barry et al. [20] used the technique of (Parlange and Brutsaert) and showed that the influence of the capillary fringe increases with oscillation frequency. Another author, Li et al. [18], developed a modified kinematic boundary condition for the water table, which takes capillarity effects into account.

---

[1] Note: a vertical flow model Zs(t) based solely on the second hypothesis is conceivable, but it would be limited to merely ensuring mass conservation between the bottom boundary and the mobile water table.

Most of the above-cited modification techniques were based in some way on the Green-Ampt (G-A) "piston flow" approximation [21]. The latter assumes that the equivalent capillary fringe is completely saturated with water, and that the sharp interface between the capillary fringe and the dry medium is characterized by a fixed suction head called "wetting front suction" and denoted $\Psi_{FRONT}$ or $\Psi_F$ [19]. Strictly speaking, the classical G-A model is limited to the case of vertical downward infiltration under constant positive pressure imposed at soil surface, although some variants have been introduced in the literature. [2] Now, the classical G-A infiltration problem and the tidal oscillations considered in this work are both vertical flow configurations, but they differ in two important aspects: (1) the time-varying boundary condition considered in this work is an oscillatory "tidal" pressure; and (2) the tidal pressure condition is imposed at the bottom boundary (not at the top) so that the forcing in our tidal problem is alternately co- and contra-gravitational (while G-A infiltration is co-gravitational at all times).

Thus, Nielsen and Perrochet [14,15] analyzed a tidal oscillation problem similar to ours, and they attempted to include capillary effects in their analyses. At first, they accounted for the effects of the capillary fringe above a vertically oscillating water table by introducing an equivalent (integral) capillary fringe height hc(t), which leads to a differential equation similar to the G-A model. They proposed to solve this equation for small amplitude using complex variables. The result of their linearized small amplitude model is expressed in terms of a (real or complex) dynamic effective porosity "$n_D$", which accounts the total equivalent saturated height in the column (water table height + equivalent thickness of the capillary fringe). The authors compared their complex effective porosity model to a laboratory experiment conducted on a vertical column, and they also introduced a small amplitude model based directly on the Green-Ampt piston flow approximation expressed in terms of front suction, which they named "Hc". In order to obtain better fit with the experiment, other modifications were introduced (e.g., an empirical hydraulic conductivity $K < K_S$ was introduced in the equivalent capillary fringe zone). Overall, the proposed analytical models are limited to small amplitude scenarios and/or contain empirical parameters. The authors [14,15] conclude that (i) the various variants of the small amplitude G-A models do not match the experimental data very well (in terms of amplitude and phase of the frequency response); (ii) the results of the "dynamic effective porosity" ($n_D$), from their equivalent integral capillary fringe model, cannot be matched simultaneously for amplitude and phase for any real value of "$n_D$", but the match becomes good with an empirically fitted complex-valued "$n_D$". Note: this boils down to fitting empirically the amplitude decay and phase shift in a G-A type solution of water table fluctuations.

---

[2] Note: anticipating on the next sections, it can be mentioned here that the multi-front models introduced in this paper are themselves extensions of the Green-Ampt approach.

Other works in the literature extended the G-A approximation in various ways, e.g. for infiltration under time varying conditions (quite different from oscillatory water table problems as reviewed above), and also, for layered soils, for multidimensional flow configurations, etc. These are briefly reviewed below.

Warrick et al. 2005 [22] used and solved a slightly modified form of the G-A infiltration equation for the case of irrigation under time-variable ponded depth. They compared the results to field observations in two irrigated plots: (a) input water depth "d" = $h_{SURF}(t)$, and (b) output cumulative infiltration I(t) (wetting front depth $Z_F(t)$ could be deduced from porosity). Their solution technique is to discretize the non linear Green-Ampt ODE with relatively coarse time steps, such that the prescribed ponded depth $h_{SURF}(t)$ is assumed constant within each step (piecewise constant). The solution of this variable time G-A infiltration was also compared to a numerical finite element solution of Richards equation with the HYDRUS1D code. The results were close in terms of infiltration depth I(t), but it should be noted (along with the authors) that the effect of time variability is not strong in this type of ponded infiltration scenario: indeed the solutions I(t) obtained with time varying ponded depth (HYDRUS1D, G-A) and with constant mean ponded depth (G-A) were all quite close. The conclusion of the authors is that time variability of ponded depth does not have a great effect on the resulting cumulative infiltration I(t).

The latter observation should be related to our previous remarks concerning co-gravity flows versus contra-gravity flows. The flow system studied by Warrick et al. 2005 [22] is co-gravity at all times (despite time variability at soil surface). In the tidal case, we have alternating co-/contra-gravity flow.

At this point, it may be useful to consider a technical mathematical issue that emerges from the above reviewed works. It concerns the "discretization" of the Green-Ampt model in various situations and for various purposes… Indeed, in the remainder of the present work, we will propose to improve on the classical G-A approach by discretizing the state variables themselves (the water contents and the suctions, and consequently the unsaturated permeability along the profile). In comparison, we have seen just above that Warrick et al. (2005) [22] propose to discretize the G-A model in time to reformulate it as a classical G-A model under piecewise constant conditions. The resulting nonlinear ODE can then be solved either by numerical integration or, say, by explicit Runge-Kutta finite differences in time[3]. On the other hand, in finite element or finite volume methods (HYDRUS1D code, BIGLOW3D code) the non linear Richards PDE (Partial Differential Equation) is discretized in both time *and* space.

Ma et al. 2010 [23] empirically adapted and discretized the G-A model for infiltration in layered soils, based on previous work which appears to have been published in a journal in chinese - mandarin language ("*Han et al. 2001, Chinese J. of EcoAgriculture*"). Their modification leads to a layer-discretized G-A infiltration equation governing the wetting front elevation $Z_F(t)$ as it passes through each layer, taking into account, in this process, the different effective saturations and effective harmonic mean conductivities as the wetting front moves downwards. Some parameters of the governing equation were determined empirically, e.g., the ratio of actual measured moisture volume divided by total saturated volume. The results obtained for a 5-layer soil compared favorably with numerical solutions of the Richards equation (HYDRUS1D code), but only in terms of infiltration rate and cumulative infiltration $I(t)$. However, tracking down the motion of the wetting front was less successful, in comparison with experimental results.

Kacimov et al. 2010 [24] consider also the case of infiltration in a vertically heterogeneous soil using the Green-Ampt approach, but they differ in two ways from the previous study: they treat the case of a continuously stratified soil, and their approach is more formal (quasi-analytical). The authors take into account a continuously varying saturated conductivity $K_S(z)$ and wetting front suction $\Psi_F(z)$, the latter being considered as a capillary parameter of the soil. This leads to a nonlinear ODE (Ordinary Differential Equation), which they end up solving numerically for the case of exponentially varying $K_S(z)$ and $\Psi_F(z)$ (using MATHEMATICA's "NDSolve" package). The authors also introduce randomness in the exponential profile $K_S(z)$ and briefly analyze the results in a probabilistic framework (mean and dispersion variance of the infiltration process).

Selker et al. 1999 [25] treat the case of infiltration into vertically stratified soils using the Green-Ampt approach and a parametrization of heterogeneity based on pore size distribution. Considering a continuously variable pore size with depth, both $K_S(z)$ and $\Psi_F(z)$ vary with depth. More precisely, these authors use a relationship between $K_S(z)$ and $\Psi_F(z)$ in terms of pore size distribution, and they assume for simplicity that $\theta s(z)$ is nearly constant based on field evidence. The resulting quasi-analytical model is applied to monotonic variation with depth (linear, power law and exponential).

---

[3] Note: in all such cases, when the resulting equations boil down to single integrals or to integrable 1rst order ODE's, we refer to these as quasi-analytical or semi-analytical solutions.

Chen and Young 2006 [28] developed a G-A infiltration model orthogonally to the surface, in the case of a sloping soil surface. Technically, their flow solution is 1D, although the flow direction is not vertical. The flow domain is semi-infinite normally to slope. The sloped soil G-A solution resembles the classical G-A solution (semi-analytical). The authors use it for studying hypothetical hydrologic scenarios of ponded infiltration. They also extend the model to treat the case of steady or transient rainfall (before and after ponding). One of their conclusions is that, under constant ponding head, the infiltration rate normal to the sloping surface is enhanced by the slope angle compared to the case of a horizontal soil. But this effect is only significant at early times when capillary effects dominate over gravity. At later times, or as t→∞, the gravitational infiltration rate does not depend on slope angle. These results have consequences on run-off over sloping soils in watershed hydrology.

Note, in the present paper, we focus on time variability and tidal oscillations, rather than heterogeneity; however, it will be interesting in future to consider combining the previously reviewed extensions of the Green-Ampt approach for stratified soils, and for sloping soils, with the oscillatory multi-front model that is developed in the remainder of this paper.

Finally, it is also worth noting that the G-A piston flow approximation has also been applied in the literature to fully multi-dimensional flow systems. For instance, the G-A approach has been used to analyze the multidimensional growth of a wetting bulb at the bottom of a falling head permeameter: see Regalado et al. 2005 [26], whose work is based on Philip 1993 [27].

*The remainder of this article is organized as follows.*

In **Section 2**, we present briefly the configuration of the oscillatory flow system to be studied analytically and numerically in the rest of the paper. It consists in a vertical soil column comprising a water table forced by an oscillating pressure at bottom. This concept is illustrated by a brief description of an actual experiment, including an instrumented soil column connected to a "tide machine" (the complete results from this set of experiments are currently being analyzed).

In **Section 3**, we present a series of successive generalizations of the Green-Ampt approach, starting with the single front model (in the first subsection) and ending up with a general multi-front or "N-front" model. The single front model is, in a sense, an "inverted" and "oscillatory" version of the classical Green-Ampt infiltration model. The 2-front and then the N-front models constitute further extensions, where the N fronts are defined based on the nonlinear pressure-dependent curves (θ(h), K(h)), i.e., respectively, water content vs. pressure and hydraulic conductivity vs. pressure. The resulting equations are nonlinear systems of 1rst order ODE's. We focus on large amplitude oscillations of the entry pressure head, up to 100% compared to the static height of the water table (most of the cases presented in this paper are based on this extreme case). The results of the 2-front and of the 20-front model are discussed, including comparison tests with numerical finite volume solution of the nonlinear Richards equation (PDE).

**Section 4** discusses in detail the parametrization of the single front, 2-front, and more generally the N-front models. For instance, the front suction $\psi_F$ is a parameter of the single front model, and it is computed from the unsaturated properties K(h) and/or θ(h) of the unsaturated soil. This concept is extended to the N suctions at the N fronts. Other parameters intervene. All parameters of the proposed multi-front model are presented and discussed in this section.

**Section 5** discusses the performance of the single front, the 2-front, and the multi-front model (the latter with N = 20) for two different soils (fine sand SilicaSand, and Guelph Loam) and for very large amplitude of the entry pressure. The models are compared to a refined solution of the Richards PDE using the implicit finite volume code BIGFLOW. The comparisons are based on water table height Zs(t) and bottom flux qo(t) (signals).

**Section 6** presents our conclusions and a brief summary of the work, and discusses various extensions and perspectives. It is pointed out that the multi-front approach is also a good approximation of the Richards equation, not only in terms of water table signals, but also in terms of water content vertical profiles (for N ≥ 10 to 20 fronts). This section also illustrates a possible application of the multi-front approach to parametric study of the frequency response to tidal forcing (behavior of the mean water table height vs. frequency for the Guelph Loam).

## 2. Oscillatory flow in 1D porous column

In this article, we consider a partially saturated 1D porous column under "dynamic" conditions. The dynamic effect is simulated by an oscillatory "entry pressure head" or "driving head" imposed at the bottom face of the column. It can be expressed as:

$$h(z=0,t) = h_0(t) = \overline{h_0} + A_0 sin(\omega_0 t) \qquad 1$$

where:
- $\overline{h_0}$ is the positive time-averaged entry pressure head, chosen to coincide with the initial hydrostatic level in the column;
- $A_0$ is the amplitude of the entry pressure;

- $\omega_0 = 2\pi / T_P$ is the angular frequency, and
- $T_P$ is the period of the imposed entry pressure.

Experimentally this driving head can be generated by a tide machine connected to the bottom of the column as described in [29]. This concept is illustrated by the schematic the schematic in **Fig. (1)**, describing an actual instrumented soil column and the associated tide machine. This experiment was conducted at the Institut de Mécanique des Fluides de Toulouse; the results of which are still being analyzed (some of these experimental results were reported in [29]).

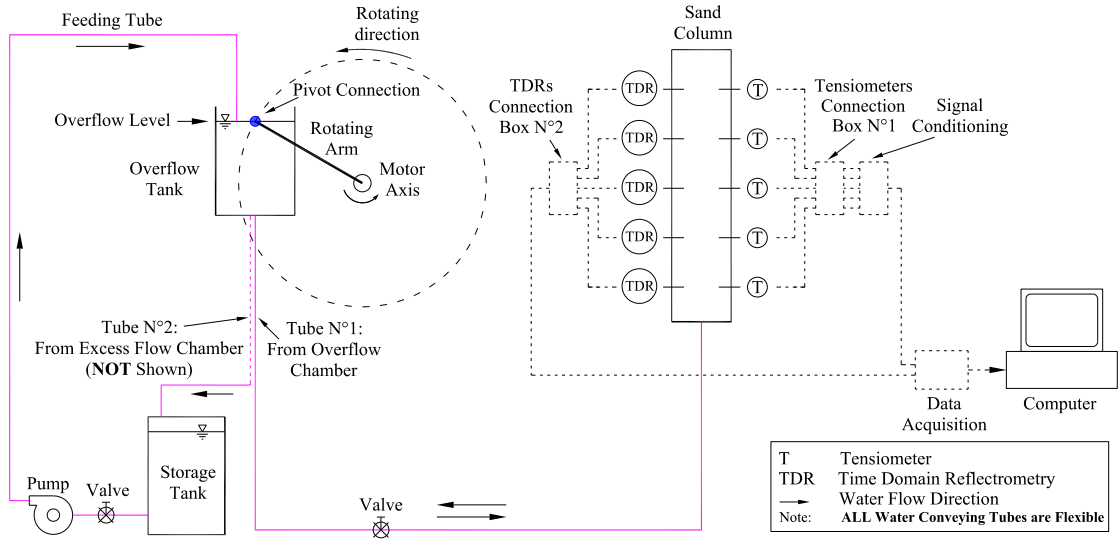

**Fig. 1:** Schematic diagram of the tide machine, soil column and measurement system.

### 3. Generalised Green-Ampt models for vertical oscillatory flows

The generalized Green-Ampt model presented in this section is applied to (and validated for) the case of partially saturated/unsaturated oscillatory flows such as occurs in beaches under tidal forcing. On the other hand, Alastal et al. [38] developed, applied, and validated this multi-front model for quite different cases of non-oscillatory flow, including vertical infiltration downwards towards water table (shallow or deep); sudden capillary rise (a water table is suddenly imposed at depth z=L in an initially dry soil); and a case of gradual water table rise. The Multi-Front results for these non-oscillatory flows were benchmarked with fixed grid unsaturated flow codes, and furthermore the order of accuracy was analyzed in terms of the number of fronts: for the case of infiltration towards a deep water table, the error on volumetric water contents (θ) decreased as M to the power (-1.92) with the number M of discrete fronts, and the corresponding exponent for the error on the surface infiltration flux was (-0.99).

### 3.1. The classical Green-Ampt model (vertical infiltration downwards)

The Green-Ampt model [21] has been the focus of many interests because of its simplicity and satisfactory performance for a variety of hydrological applications [23,30]. It was originally developed to study one dimensional (1D) vertical infiltration into homogeneous porous medium of infinite depth. The Green-Ampt infiltration model assumes piston flow, with a sharp wetting front that separates the saturated zone from the unsaturated zone. This wetting front is characterized by a constant suction head (considered a parameter of the soil), and it propagates downward into the soil under the combined action of capillary pressure gradient and gravity. Initially, the soil is assumed to have a constant, uniform water content profile.

As we have seen earlier, a number of studies have focused on extending the Green-Ampt model to other applications. In this article, the Green-Ampt (G-A) model is generalized to deal with 1D oscillatory flow in partially saturated/unsaturated porous media, taking into account unsaturated capillary effects on the oscillation process as realistically as possible. In fact, several generalized versions of the G-A model are developed for oscillatory flow, starting below with the basic single front model, and ending up with the general multi-front or N-front model.

### 3.2. The Green-Ampt or single front model, extended to oscillatory flows

The purpose of this first basic model is to provide a semi-analytical solution for water table height fluctuations, and also, to provide a basis for further extensions (multi-front). This model is basically an "inverted" version of the classical Green-Ampt infiltration model. A pressure condition is imposed at bottom (instead of top), and the model is further extended to accommodate periodic fluctuations of the imposed bottom pressure. Thus, strictly speaking, the so-called "wetting front" is in fact either wetting or draining, depending on time. This model is also called "single front" model because, as in the classical G-A approach, it is based on the movement of a single front ($Z_F(t)$).

Accordingly, based on the usual Green-Ampt piston flow hypotheses, we assume that:
- There exists a well defined "wetting front" separating the fully saturated and the totally dry regions, as shown in **Fig. (2),** where the dashed line represents the GA approximation.
- The wetting front is assumed to be characterized by some effective, constant suction head $\psi_F$, or pressure head $h_F$, with $\psi_F = -h_F$.

The free surface (water table) is itself characterized by h = 0 (zero pressure head relative to air pressure). Both the wetting front elevation, $Z_F(t)$, and the free surface elevation, $Z_S(t)$, move in response to the oscillating bottom pressure head.

Note that the entry pressure head is assumed in this work to be a simple harmonic (sine or cosine function); the response of the wetting front and water table may also be analyzed with the same techniques when the forcing is a more complex signal containing, say, both discrete and continuous spectra (not just a simple harmonic).

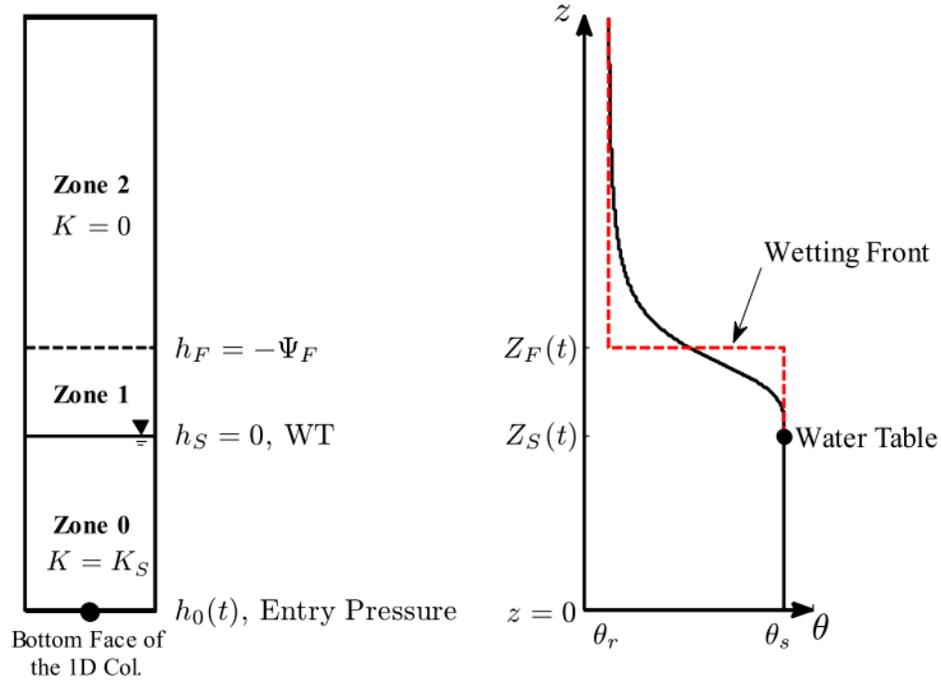

**Fig. 2:** schematic diagram of the single front model. On the left: 1D porous column shows the water table ($Z_S(t)$) and the wetting front ($Z_F(t)$) elevations where the suction heads are 0 and $\psi_F$ respectively. On the right: the instantaneous water content profile with Green-Ampt approximation (dashed line)

Our single-front model approximation divides the soil column into three zones, labeled i = {0,1,2} from bottom to top, each zone having *a priori* its own hydraulic conductivity and water content (Ki, θi):

- **Zone 0**: this zone is totally saturated; it extends from the bottom of the column ($z = 0$) to the moving water table ($z = Z_S(t)$); in this zone, $K_0 = K_S$, $\theta_0 = \theta_S$.
- **Zone 1**: this zone extends from the moving water table ($z = Z_S(t)$) to the moving wetting front ($z = Z_F(t)$); it represents a wet or quasi-saturated transition zone between the totally saturated Zone 0 and the totally dry Zone 2; the properties of this zone are $K_1 = K_S$ and $\theta_1 = \theta_S$, although lower values can be assumed empirically (say $K_1 \leq K_S$).

- **Zone 2**: this last zone is the totally dry region extending from the wetting front ($z = Z_F(t)$) up to infinity; it is a semi-infinite zone; the flux is null in this zone, since we choose $K_2 = 0$ and $\theta_2 = \theta_r$.

Let us now develop the governing equations under these assumptions, using mass conservation principles and Darcy's flux-gradient law (a simplified version of Navier-Stokes' conservation of momentum).

### 3.3. Darcy's flux-gradient equation for the single front model

Now, let us apply Darcy's equation to calculate the vertical fluxes q(z,t).

The flux $q_0(t)$ in the saturated *Zone 0* is constant in space, and is given by:

$$q_0(t) = -K_0 \left( \frac{\partial h}{\partial z} - g \right) = -K_S \left( \frac{0 - h_0(t)}{Z_S(t)} - g \right) = K_S \left( \frac{h_0(t)}{Z_S(t)} + g \right) \qquad 2$$

where $Z_S(t)$ is the water table elevation; $h_0(t)$ is the bottom entry pressure head. Note that "g" represents unit gravity, with $g = -1$ if the Oz axis is directed upwards (as is the case here). In fact, the solution is also generalized in terms of the gravity vector $g$:

- $g = -1$ in the present case (pressure fluctuations at bottom, $z$ directed upwards).
- $g = +1$, e.g., for variable-head ponded infiltration, with $z$ directed downwards.
- $g = 0$, e.g. for horizontal flow model and/or in the absence of gravity.

The flux $q_1(t)$ in the quasi-saturated transition *Zone 1* is also constant in space, given by:

$$q_1(t) = -K_1 \left( \frac{h_F - 0}{Z_F(t) - Z_S(t)} - g \right) = K_S \left( \frac{\psi_F}{Z_F(t) - Z_S(t)} + g \right) \qquad 3$$

where $h_F = -\psi_F$ is the constant suction head at the sharp wetting front (it should be considered a parameter characterizing, mainly, the unsaturated soil properties).

### 3.4. Mass conservation equations for the single front model

On the other hand, let us apply the mass conservation equation locally around the moving water table ($z = Z_S(t)$). This yields:

$$q_0(t) - q_1(t) = (\theta_0 - \theta_1) \cdot \frac{dZ_S}{dt} \qquad 4$$

where $\theta_0$ and $\theta_1$ are the water content in the zone 0 and 1 respectively. However, both of these equal the saturated water content ($\theta_S$) based on the Green–Ampt assumptions. Therefore for this simple single front model:

$$q_0(t) = q_1(t) \qquad 5$$

Now, applying the mass conservation equation through the moving wetting front ($z = Z_F(t)$):

$$q_1(t) - q_2(t) = (\theta_1 - \theta_2) \cdot \frac{dZ_F}{dt} \quad\quad 6$$

where $q_2$ and $\theta_2$ are the flux and water content in the dry zone 2 ($q_2 = 0$, $\theta_2 = \theta_r$); and $\theta_1 = \theta_S$. Then:

$$q_1(t) = (\theta_S - \theta_r) \cdot \frac{dZ_F}{dt} \quad\quad 7$$

Combining Eq. (3), (5) and (7):

$$\frac{dZ_F}{dt} = \frac{K_S}{\theta_S - \theta_r} \left( \frac{\psi_F + h_0(t)}{Z_F(t)} + g \right) \quad\quad 8$$

Eq. (8) is an ordinary differential equation (ODE) that can be solved in the time domain for our assumed initial condition: $Z_F(t=0) = \overline{h_0} + \psi_F$. This equation, together with the initial condition, is solved for the unknown wetting front height as a function of time using the ODE MATLAB solver package ("ODE15S" stiff solver was used as in Alastal al. [38], or alternatively, we have also used here "ODE23T" stiff solver in some cases).

Once $Z_F(t)$ is determined, Eq. (5) is solved to obtain the water table height $Z_S(t)$:

$$Z_S(t) = Z_F(t) \left( \frac{h_0(t)}{h_0(t) + \psi_F} \right) \quad\quad 9$$

Finally, either Eq. (2) or Eq. (3) may be applied to calculate the bottom flux [$q_0(t)$ or $q_1(t)$].

### 3.5. The two-front model

In this section, the novel two-front model is considered. The water content is approximated by two sharp fronts instead of one. Our expectation is to modify the results by approximate the characteristic curves by two fronts instead of only one, **Fig. (3)**. Each front represents an equipotential line (of suction head) that moves with time as a response to the bottom entry fluctuating pressure head.

The two-front model approximation divides the column into four regions:

- **Zone 0**: it is a totally saturated zone extending from the bottom of the column ($z=0$) to the moving water table ($z = Z_S(t)$); with $K_0 = K_S$, $\theta_0 = \theta_S$
- **Zones 1** and **2**: They are comprised between the moving water table ($z = Z_S(t)$) and the second moving wetting front ($z = Z_{F2}(t)$). These zones can be considered as "stepped" transition zones from the totally saturated zone to the totally dry zone. The parameters are: $K_1 = K_S$ and $K_2 < K_S$; $\theta_1 = \theta_S$ and $\theta_2 < \theta_S$.

- **Zone 3**: This is the totally dry region, it is a semi-infinite zone starting above the second moving front ($z = Z_{F2}(t)$); the flux is null in this zone, and the parameters are: $K_3 = 0$; $\theta_3 = \theta_r$.

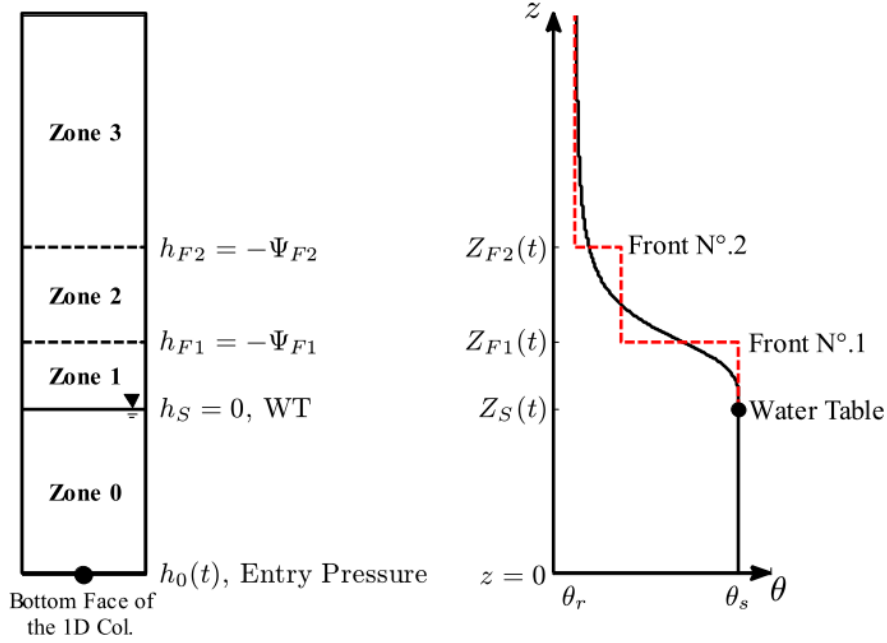

**Fig. 3**: Schematic diagram of the two-front model approximation. On the left: 1D porous column shows the water table ($Z_s(t)$) and the two fronts ($Z_{F1}(t), Z_{F2}(t)$) elevations where the suction heads are 0, $\psi_{F1}$ and $\psi_{F2}$ respectively. On the right: the instantaneous water content profile with two-front model approximation (dashed line).

### 3.6. Darcy equations for the two front model

Now, let us apply Darcy's law to determine the flux in each zone of the two front model.

- Flux in zone 0, $q_0(t)$

$$q_0(t) = -K_0 \left( \frac{\partial h}{\partial z} - g \right) = -K_S \left( \frac{0 - h_0(t)}{Z_S(t)} - g \right) = K_S \left( \frac{h_0(t)}{Z_S(t)} + g \right) \qquad 10$$

- Flux in zone 1, $q_1(t)$

$$q_1(t) = -K_1 \left( \frac{\partial h}{\partial z} - g \right) = -K_1 \left( \frac{h_{F1} - 0}{Z_{F1}(t) - Z_S(t)} - g \right) = K_1 \left( \frac{\psi_{F1}}{Z_{F1}(t) - Z_S(t)} + g \right) \qquad 11$$

- Flux in zone 2, $q_2(t)$

$$q_2(t) = -K_2\left(\frac{\partial h}{\partial z} - g\right) = -K_2\left(\frac{h_{F2} - h_{F1}}{Z_{F2}(t) - Z_{F1}(t)} - g\right) = K_2\left(\frac{\psi_{F2} - \psi_{F1}}{Z_{F2}(t) - Z_{F1}(t)} + g\right) \quad 12$$

The flux is null in the dry zone 3 ($q_3(t) = 0$).

### 3.7. Mass Conservation equations for the two front model

On the other hand, let us apply the mass conservation equation locally around each moving front (including the water table).

- Mass conservation equation traversing the moving water table ($z = Z_S(t)$):

$$q_0(t) - q_1(t) = (\theta_S - \theta_1) \cdot \frac{dZ_S}{dt} \quad 13$$

- Mass conservation equation traversing the first moving front ($z = Z_{F1}(t)$):

$$q_1(t) - q_2(t) = (\theta_1 - \theta_2) \cdot \frac{dZ_{F1}}{dt} \quad 14$$

- Mass conservation equation traversing the second moving front ($z = Z_{F2}(t)$):

$$q_2(t) - 0 = (\theta_2 - \theta_r) \cdot \frac{dZ_{F2}}{dt} \quad 15$$

Eqs. (13), (14) and (15) can be written in matrix form as follow:

$$\begin{bmatrix} \theta_S - \theta_1 & 0 & 0 \\ 0 & \theta_1 - \theta_2 & 0 \\ 0 & 0 & \theta_2 - \theta_r \end{bmatrix} \begin{Bmatrix} dZ_S/dt \\ dZ_{F1}/dt \\ dZ_{F2}/dt \end{Bmatrix} = \begin{Bmatrix} q_0(t) - q_1(t) \\ q_1(t) - q_2(t) \\ q_2(t) \end{Bmatrix} \quad 16$$

where $\theta_1$ is usually equal to $\theta_S$ (therefore the "mass matrix" on the left hand side is usually singular). This is a system of differential algebraic equations (DAE) that we solve using the ODE MATLAB package "ODE15S" based on multistep finite difference schemes, adapted for stiff non linear algebraic differential systems having possibly a singular mass matrix (alternatively, we have also used "ODE23T"). The system is subjected to the following initial condition:

$$\begin{cases} Z_S(t=0) = \overline{h_0} \\ Z_{F1}(t=0) = \overline{h_0} + \psi_{F1} \\ Z_{F1}(t=0) = \overline{h_0} + \psi_{F2} \end{cases} \quad 17$$

Once the evolution of the water table height ($Z_S(t)$) and front elevations ($Z_{F1}(t)$, $Z_{F2}(t)$) were determined; then eq. (10), (11) and (12) can be applied to calculate the fluxes.

## 3.8. The multi-front model

In this section, we generalize the inverse Green-Ampt model to the multi-front model. The unsaturated profiles, K(z) and θ(z), are approximated by *N* fronts as shown **Fig. 4**. With this parametrization of the multi-front model, we expect that the accuracy of the results will be improved as the transition zone is divided (approximated) by a larger number of "fronts" over the height of the 1D column. As a consequence, the profiles K(z) and θ(z) will also be more refined.

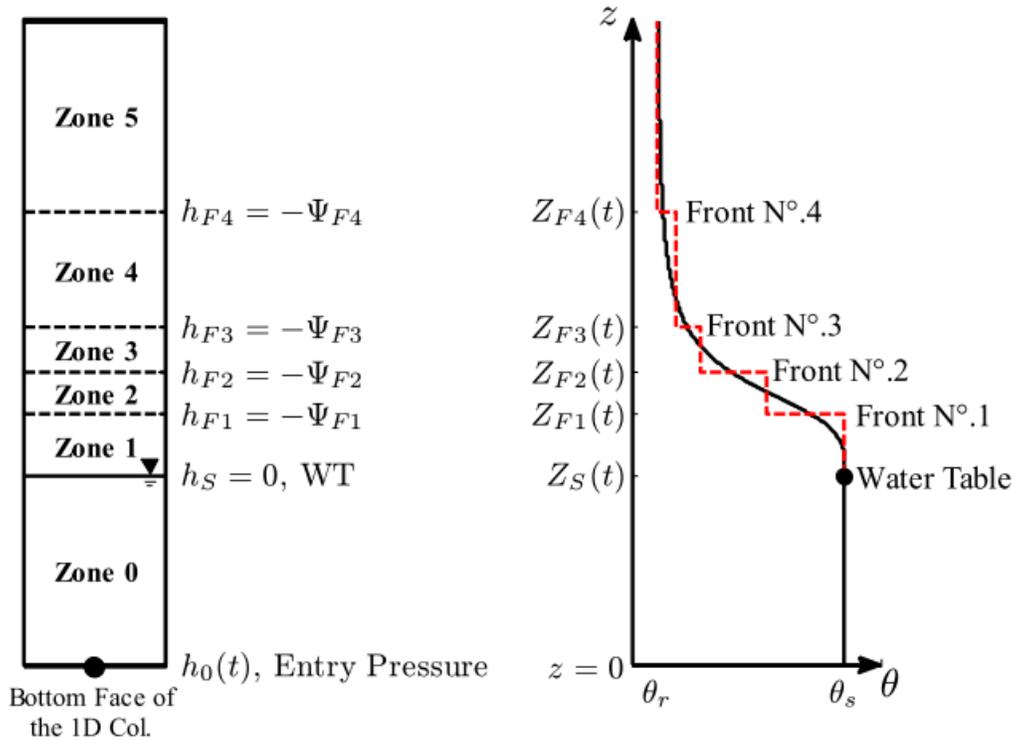

**Fig. 4:** Schematic diagram of a multi-front model approximation with four fronts (for clarity we show only 4 fronts). On the left: 1D porous column shows the water table ($Z_s(t)$) and the four successive fronts at elevations of ($Z_{F1}(t)$, $Z_{F2}(t)$, $Z_{F3}(t)$ and $Z_{F4}(t)$) where the suction heads are 0, $\psi_{F1}$, $\psi_{F2}$, $\psi_{F3}$ and $\psi_{F4}$ respectively. On the right: the instantaneous water content profile with multi-front model approximation (dashed line).

The multi-front approximation divides the 1D column into N+2 regions
- **Zone 0**: it is a totally saturated zone extending from the bottom of the column ($z=0$) to the moving water table ($z=Z_s(t)$); parameters: $K_0 = K_s$, $\theta_0 = \theta_s$

- **Zones 1 to N**: These zones are comprised between the moving water table ($z = Z_S(t)$) and the last (Nth) moving front ($z = Z_{F_N}(t)$). They are the N transition zones from the totally saturated zone to the totally dry one. The parameters of each zone (number j) are Kj ≤ Ks and θj ≤ θs.
- **Zone N+1**: This is the totally dry region; it is a semi-infinite dry zone above the $N^{th}$ moving front ($z = Z_{F_N}(t)$); the flux in this zone is null, and the parameters are: $K_{N+1} = 0$, $\theta_{N+1} = \theta_R$.

## 3.9. N-front model: applying Darcy's equation for each zone

- Applying Darcys law for Zone 0

$$q_0(t) = K_S \left( \frac{h_0(t)}{Z_S(t)} + g \right) \quad\quad 18$$

- In the same manner the flux in the zone 1:

$$q_1(t) = K_1 \left( \frac{\psi_{F1}}{Z_{F1}(t) - Z_S(t)} + g \right) \quad\quad 19$$

- For the remaining zones (zone i, $2 \leq i \leq N$)

$$q_i(t) = K_i \left( \frac{\psi_{F_i} - \psi_{F_{i-1}}}{Z_{F_i}(t) - Z_{F_{i-1}}(t)} + g \right) \quad\quad 20$$

- For the top dry zone (zone N+1): $q_{N+1}(t) = 0$

## 3.10. N-front model: applying mass conservation through the fronts

Moreover, applying the mass conservation equation through the moving water table, and through the other N fronts (interfaces), we obtain the following results.

- Mass conservation equation traversing the moving water table ($z = Z_S(t)$):

$$q_0(t) - q_1(t) = (\theta_S - \theta_1) \cdot \frac{dZ_S}{dt} \quad\quad 21$$

- Mass conservation equation traversing the i-th moving front ($1 \leq i \leq N-1$):

$$q_i(t) - q_{i+1}(t) = (\theta_i - \theta_{i+1}) \cdot \frac{dZ_{Fi}}{dt} \quad\quad 22$$

- Mass conservation equation traversing the last (Nth) moving front ($z = Z_{F_N}(t)$):

$$q_N(t) - 0 = (\theta_N - \theta_r) \cdot \frac{dZ_{F_N}}{dt} \quad\quad 23$$

The above equations can be arranged in the following matrix form:

$$\begin{bmatrix} \theta_S - \theta_1 & 0 & \cdots & 0 & 0 \\ 0 & \theta_1 - \theta_2 & \cdots & 0 & 0 \\ \vdots & \vdots & \ddots & \vdots & \vdots \\ 0 & 0 & \cdots & \theta_{N-1} - \theta_N & 0 \\ 0 & 0 & \cdots & 0 & \theta_N - \theta_r \end{bmatrix} \begin{Bmatrix} dZ_S/dt \\ dZ_{F_1}/dt \\ \vdots \\ dZ_{F_{N-1}}/dt \\ dZ_{F_N}/dt \end{Bmatrix} = \begin{Bmatrix} q_0(t) - q_1(t) \\ q_1(t) - q_2(t) \\ \vdots \\ q_{N-1}(t) - q_N(t) \\ q_N(t) \end{Bmatrix} \qquad 24$$

This is a system of differential algebraic equations (DAE) that can be solved by the ODE package of MATLAB (ODE15S) subject to the following initial condition:

$$\begin{cases} Z_S(t=0) = \overline{h_0} \\ Z_{F_1}(t=0) = \overline{h_0} + \psi_{F_1} \\ \vdots \\ Z_{F_N}(t=0) = \overline{h_0} + \psi_{F_N} \end{cases} \qquad 25$$

Once the evolution of the water table height ($Z_S(t)$) and the fronts elevations ($Z_{F_1}(t)$, $Z_{F_2}(t)$ to $Z_{F_N}$) are determined; then eq. (21), (22) and (23) can be applied to calculate the fluxes between the moving fronts, including also the bottom flux ($q_0(t)$).

## 4. Generalized Green-Ampt models parameters

The performance of the proposed models depends largely on the suitable estimation of the model parameters. These parameters are the constant suction head ($\psi_F$) at each front, the hydraulic conductivity ($K$) and the water content ($\theta$) between the successive fronts.

### 4.1. The single front model parameters

The single front model parameter [the front suction ($\psi_F$)] can be related to soil hydraulic characteristics. Two expressions for $\psi_F$ were suggested as follows:

- $\psi_F$ corresponds to the point of the inflection of the water retention curve [$\theta(h)$] at which the specific moisture capacity [$C(h) = \partial \theta(h)/\partial h$] is maximum as shown in **Fig. (5)**. The suction head at this point can be treated as a global capillary length scale of the porous media which is defined by [31]:

$$\psi_F = \frac{1}{\alpha}\left(1 - \frac{1}{n}\right)^{\frac{1}{n}} \qquad 26$$

Where $\alpha$ and $n$ are the van Genuchten model parameters.

- Another estimation of the wetting front suction is altered form Neuman [32] suggestion of the wetting front suction for the 1D infiltration:

$$\psi_F = \int_0^\infty K_r(\psi)\,d\psi \qquad 27$$

This approximation is shown in **Fig. (6)**; it is also called the Bouwer "critical suction". See Alastal et al. [38] for more details on the parametrisation of front suction.

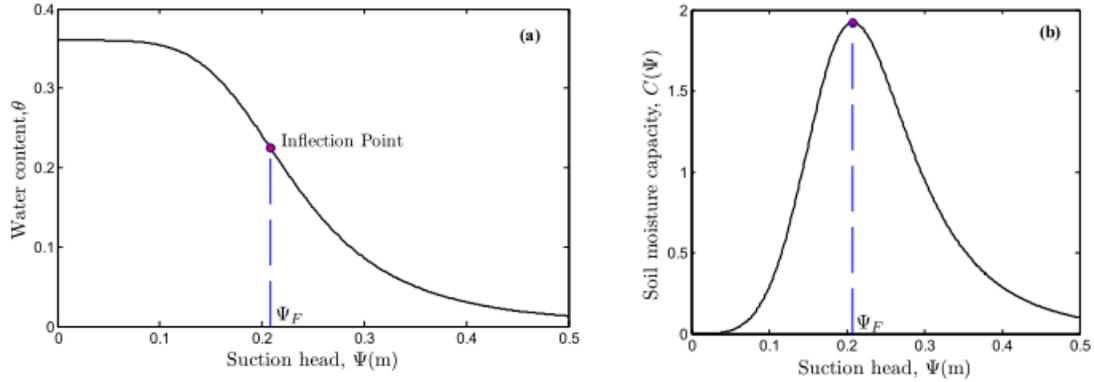

**Fig. 5:** (**a**) Water retention curve showing the point of inflection; (**b**) The corresponding to the maximum soil moisture capacity.

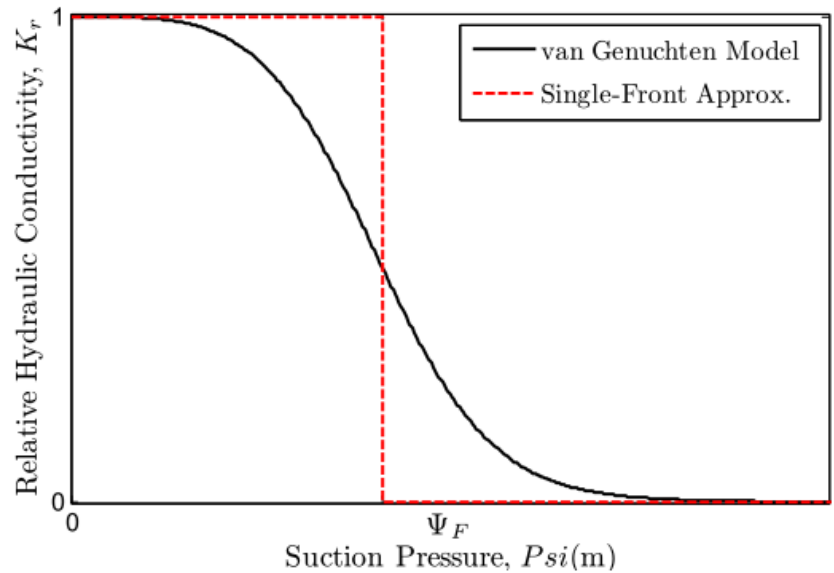

**Fig. 6:** Relative hydraulic conductivity curve with the Single-front model approximation.

## 4.2. The two-front model parameters

Here, we suggest a theoretical expression to obtain the parameters of the two-front model. The proposed method relates the two-front suction heads $\psi_{F1}$ and $\psi_{F2}$ to the soil characteristics. It is based on maintaining the area under the relative hydraulic conductivity curve $K_r(\psi)$ given by eq. (27) above and rewritten here $\left[\int_0^\infty K_r(\psi)d\psi\right]$.

In the same manner, for the two-front model, we assume that:

- There is an intermediate suction head ($\psi_{Inter}$) between the two-front suction heads ($\psi_{F1}$) and ($\psi_{F2}$) as shown in **Fig. (7)**. A good approximation of $\psi_{Inter}$ can be taken to be equal to $\psi_F$ for the single front model [see eq. (27)]. However, we can take any other value.
- The area under the $K_r(\psi)$ curve equals the area under the two-front approximation. This assumption is valid for the wall suction range [0 to $\infty$] and also for the sub regions: from [0 to $\psi_{Inter}$] and from [$\psi_{Inter}$ to $\infty$]

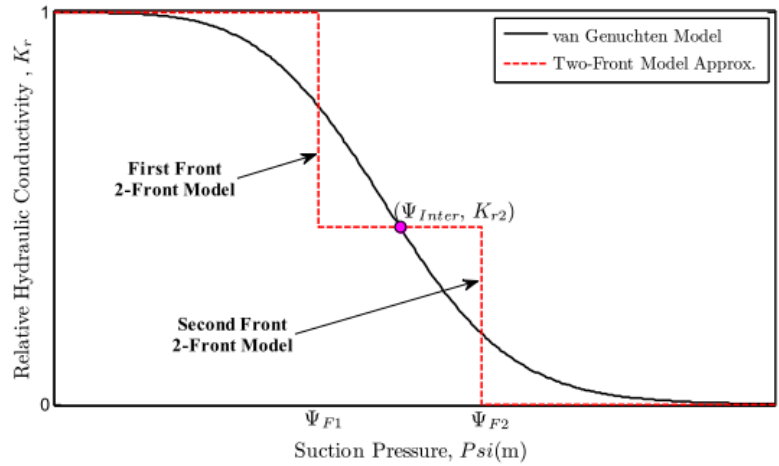

**Fig. 7:** Relative hydraulic conductivity curve with the two-front model approximation.

Thus, applying the second assumption between [0 to $\psi_{Inter}$] to obtain $\psi_{F1}$:

$$\int_0^{\psi_{Inter}} K_r(\psi)d\psi = \psi_{F1}\cdot 1 + (\psi_{Inter} - \psi_{F1})\cdot K_{r2} \qquad 28$$

And applying the second assumption between [$\psi_{Inter}$ to $\infty$] to obtain $\psi_{F2}$:

$$\int_{\psi_{Inter}}^\infty K_r(\psi)d\psi = (\psi_{F2} - \psi_{Inter})\cdot K_{r2} \qquad 29$$

Where $K_{r2} = K_r(\psi_{Inter})$ and $\psi_{Inter}$ is an intermediate value between $\psi_{F1}$ and $\psi_{F2}$.

Note that the value of the hydraulic conductivity and the water content are:

- $K_S$ and $\theta_S$ below the first front [between $Z_S(t)$ and $Z_{F1}(t)$].
- $K(\psi_{Inter})$ and $\theta(\psi_{Inter})$ between the first and second front [between $Z_{F1}(t)$ and $Z_{F2}(t)$] as shown explicitly in **Fig. (7)**.

### 4.3. The multi-front model parameters

The parameters of the multi-front model are obtained by a similar method to the one used for the two-front model parameters estimation. For $N$ fronts, the estimation of the model parameters can be summarized as follows [see **Fig.(8)**]:

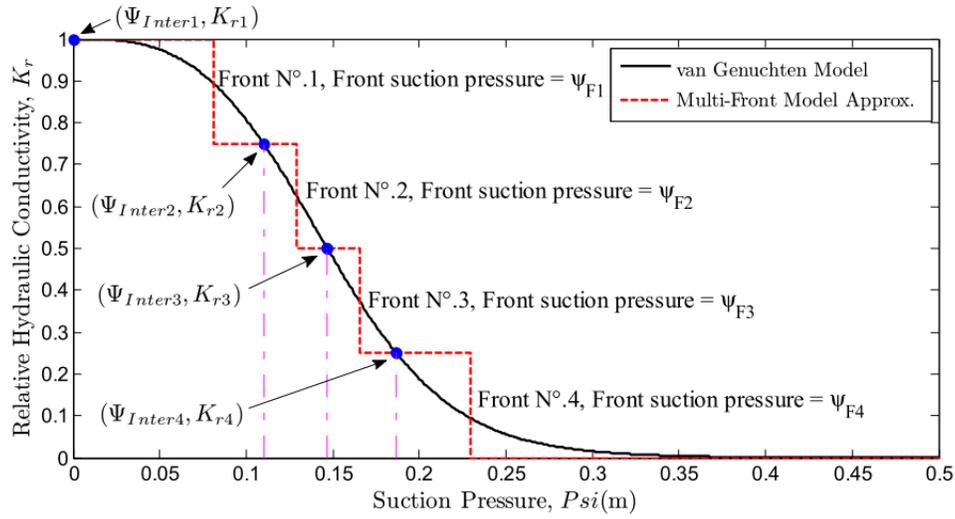

**Fig. 8:** Relative hydraulic conductivity curve with the multi-front model approximation. Four fronts are shown (to clarify the figure).

- Discretize the relative hydraulic conductivity curve $K_r(\psi)$ by $N$ intermediate suction values [$\psi_{Inter1}$ to $\psi_{InterN}$]. Where $\psi_{Inter1} = 0$.
- Calculate the relative hydraulic conductivities $K_{r_i}$ at the intermediate suctions. [$K_{r_1} = K_r(\psi_{Inter1})$ to $K_{r_N} = K_r(\psi_{InterN})$]
- Calculate the suction head at each front [$\psi_{F1}$ to $\psi_{FN}$] by keeping the area under the $K_r(\psi)$ curve equals the area under the model approximation. This step is repeated between each successive intermediate suction values. Thus, to obtain the suction front $\psi_{F_i}$ of the front number $i$, where $i=1,2,\ldots,N-1$, we have:

$$\int_{\psi_{Inter_i}}^{\psi_{Inter_{i+1}}} K_r(\psi) d\psi = \left(\psi_{Inter_{i+1}} - \psi_{Inter_i}\right) \cdot K_{r_{i+1}} + \left(K_{r_i} - K_{r_{i+1}}\right) \cdot \left(\psi_{F_i} - \psi_{Inter_i}\right) \qquad 30$$

- For the last front $\psi_{F_N}$

$$\int_{\psi_{Inter_N}}^{\infty} K_r(\psi) d\psi = \left(K_{r_N}\right) \cdot \left(\psi_{F_N} - \psi_{Inter_N}\right) \qquad 31$$

Note that the value of the hydraulic conductivity and the water content are:
- $K_S$ and $\theta_S$ between the water table and the first front [between $Z_S(t)$ and $Z_{F1}(t)$].
- $K(\psi_{Inter_i})$ and $\theta(\psi_{Inter_i})$ between front number ($i$) and front number ($i-1$), where $i = 2, 3, \ldots, N$ as shown in **Fig. (8)**.

It is worth mentioning that other method to discretize the multi-front model can be used such as: discretize the water content [$\theta(\psi)$] into N equally space segments or discretize directly the suction head $(\psi)$ ……

## 5. Validation tests, comparisons and performance of single & multi-front models

To evaluate the proposed semi-analytical Multi-Front model, the Multi-Front solutions are compared to other numerical solutions of reference, for partially saturated/unsaturated *oscillatory flows* (as occurs in beaches influenced by tides). Here, the "reference solutions" are represented by the numerical solution of the Richards equation using the Finite Volumes BIGFLOW 3D code briefly described below (section 5.1).

Two different soils, namely fine sand (FS) and Guelph loam (GL), were used in this study to evaluate the performance of the three models. The fine sand and the loam are from alastal et. al, [33]. The soil water retention curves are presented in **Fig. (9)**. The curves indicated that the fine sand has a sharp transition from the saturated to the dry water content (Green-Ampt soil behavior). In contrast with the Guelph loam which have gradual transition behavior.

The soil hydrodynamic parameters are summarized in **Table 1**; the unsaturated hydrodynamic properties of the soils are described with the soil water retention model of van Genuchten [34] (1976) in combination with the functional model of hydraulic conductivity model proposed by Mualem [35] (1976).

The validation of each of the models proposed in this study (the Green-Ampt based multi-front models) is done by comparing the evolution of water table height ($Z_S(t)$) and bottom flux ($q_0(t)$) obtained from these models, to those obtained by a refined numerical simulation based on fully implicit finite volumes with a regular mesh (BIGFLOW 3D).

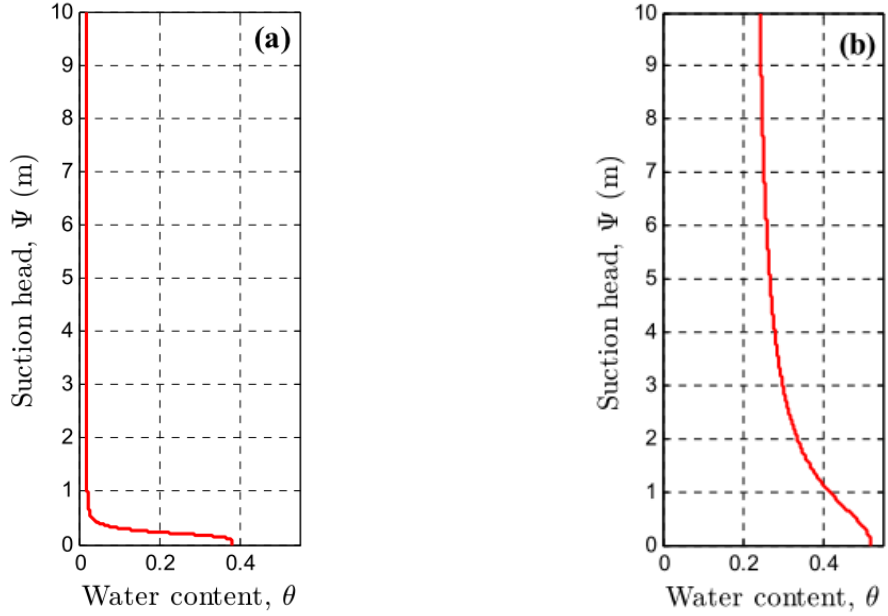

**Fig. 9:** Water retention curves for the used porous media. (a) Fine Sand "SilicaSand"; (b) Guelph Loam.

**Table 1:** Hydraulic parameters of porous media used in the study

| Parameters | Fine sand "SilicaSand" [33] | Guelph Loam [33] |
|---|---|---|
| $K_s$ (m/s) | $1.5 \times 10^{-4}$ | $3.66 \times 10^{-6}$ |
| $\theta_s$ (m³/m³) | 0.38 | 0.52 |
| $\theta_r$ (m³/m³) | 0.02 | 0.218 |
| $\alpha$ (m⁻¹) | 4.6 | 1.15 |
| $n$ | 5 | 2.03 |

The spatial gridding vertically involves 150 to 200 nodes or more, depending on the case tested. It should be noted that the multi-front or generalized G-A solutions are computed in a semi-infinite vertical domain, whereas the finite volume solution requires a finite column length with either dry suction or zero flux condition at the top. Here the chosen boundary condition was zero flux at the top of the column (preliminary tests were performed with the Richards / Bigflow solver to ensure that the results were not sensitive to column height, i.e., that the numerical column was long enough to emulate the case of a semi-finite column).

For all cases, the tidal effect is simulated by an oscillatory "entry pressure head" of the form given in eq. (1) and re-written here: $h_0(t) = \overline{h_0} + A_0 \sin(\omega_0 t)$ with the following parameters: $\overline{h_0} = 0.5$m, $A_0 = 0.5$m, $\omega_0 = 2pi/T_p$ with a period $T_p$ of a 10 min (600sec). Note that the amplitude was chosen to be maximum ($A_0/\overline{h_0} = 1$) to examine an extreme case.

All columns were considered homogeneous, isotropic, and of semi-infinite height with $\overline{h_0}$ chosen to coincide with the initial hydrostatic level.

## 5.1. Numerical code description

Numerical simulations are conducted using the Bigflow 3D finite volume flow code that has been widely documented, tested and benchmarked [36,37]. Bigflow is based on generalized Darcy-type equation, with a mixed formulation of mass conservation, capable of simulating various types of flows within the same domain. It is of the form:

$$\begin{cases} \dfrac{\partial \theta(h,\vec{x})}{\partial t} = -\vec{\nabla}² \; \vec{q} \\ \vec{q} = -\vec{K}\left(h, \vec{\nabla} H, \vec{x}\right) \vec{\nabla} H \\ H = h + \vec{g}(\vec{x})² \; \vec{x} \end{cases} \qquad 32$$

where only the first equation is actually solved (after insertion of the second and third equations). The first equation expresses mass conservation with a known water retention curve $\theta(h)$; the second equation is a generalized nonlinear flux-gradient law with tensorial hydraulic conductivity/transmissivity ($K$); and the third equation is the relation between total head ($H$) and pressure head or water depth ($h$) via a normalized gravitational vector ($g$).

In the case of 3D flow in partially saturated/unsaturated media, '$h$' is the pressure head relative to atmospheric pressure [L], $K(h)$ is the unsaturated conductivity [$LT^{-1}$], and $\theta(h)$ is volumetric water content [$L^3/L^3$].

BIGFLOW 3D can consider any functional unsaturated nonlinear model. In this article, the Van Genuchten / Mualem (VGM) model is used to describe the constitutive relationships [$\theta(h)$, $K(h)$] as follows.

The water retention function is:

$$\frac{\theta(h) - \theta_r}{\theta_s - \theta_r} = \frac{1}{\left[1 + (-\alpha h)^n\right]^m} \qquad 33$$

where $\theta_r$ is the residual water content; $\theta_s$ is the saturated water content; $\alpha$, $n$ are the model shape parameters corresponding to average capillary length and the pore size distribution respectively ; and with $m$ related to $n$ by $m=1-(1/n)$. The corresponding function for the unsaturated hydraulic conductivity is:

$$K(h) = K_s \frac{1}{\left(1+(-\alpha h)^n\right)^{m/2}} \left(1-\left[1-\frac{1}{\left(1+(-\alpha h)^n\right)}\right]^m\right)^2 \qquad 34$$

where $K_s$ is the saturated hydraulic conductivity. The remaining variables and parameters were defined earlier.

### 5.2. Performance of the single front model

**Fig. (10)** and **(11)** illustrate the comparison between the exact solution (obtained by the numerical code) and the single front model for the fine sand and the Guelph loam respectively.

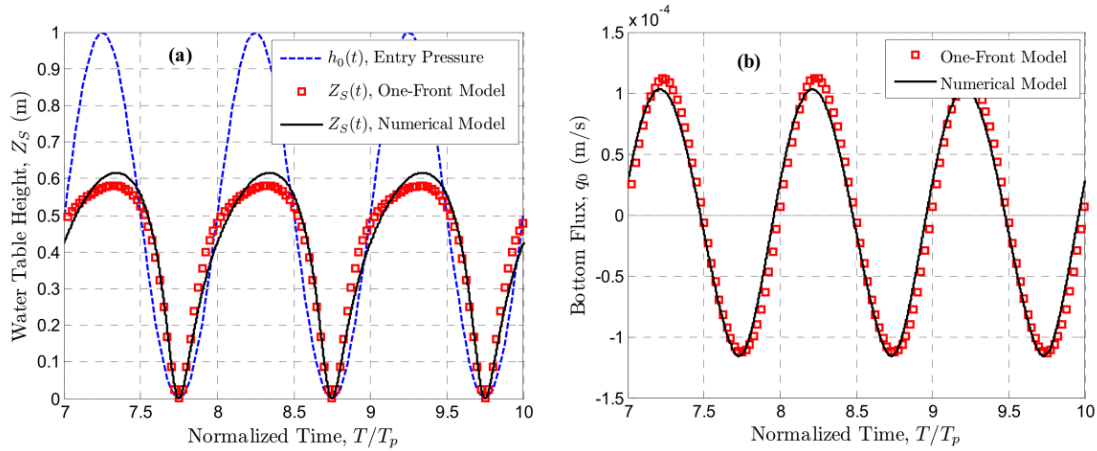

**Fig. 10:** Tidal oscillations in the SilicaSand: single front model vs. Richards. (a) Evolution of the water table height $Z_S(t)$; (b) Evolution of the bottom flux $q_0(t)$. Bold line: Quasi-exact output signals obtained from a refined numerical solution of Richards equation; squares symbols: semi-analytical solution of single-front model. The parameters of the entry bottom pressure $h_0(t)$ are: $A_0$ =0.5m, $\overline{h_0}$ =0.5m and $T_p$ =600s. The suction head at the wetting front is calculated by eq. 27 in the text.

In the case of the fine sand, the single front model was reasonably close to the exact solution whether accounting for the evolution of the water table or the bottom flux. On the other hand, for the Guelph loam the single front model fails to follow the true solution as shown in **Fig. (11)**.

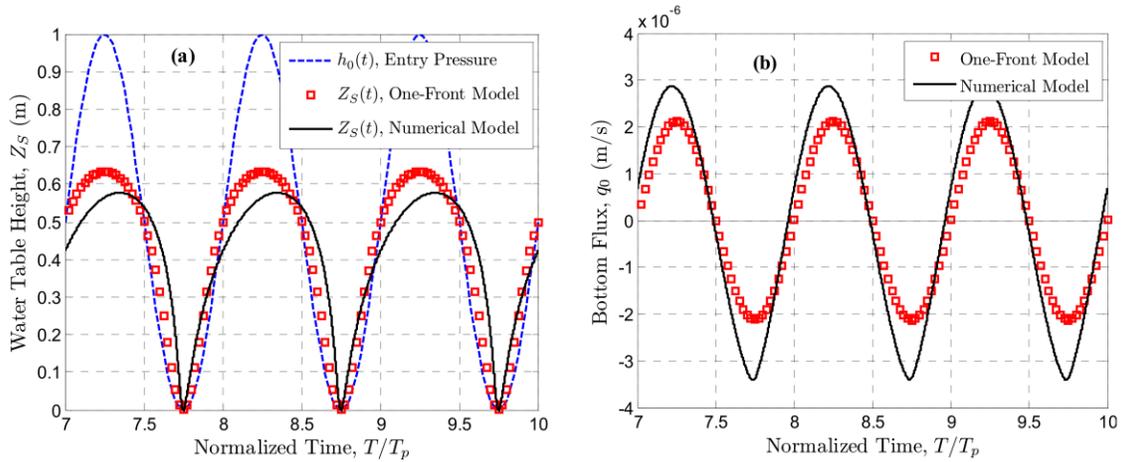

**Fig. 11:** Tidal oscillations in the Guelph Loam: single front model vs. Richards. (a) Evolution of the water table height $Z_S(t)$; (b) Evolution of the bottom flux $q_0(t)$. Bold line: quasi-exact output signals obtained from a refined numerical solution of Richards equation; squares symbols: semi-analytical solution of single-front model. The parameters of the entry bottom pressure $h_0(t)$ are: $A_0$ =0.5m, $\overline{h_0}$ =0.5m and $T_p$ =600s. The suction head at the wetting front is calculated by eq. 27 in the text.

These results were expected. The water retention curve for the fine sand has a small transition between the saturated and the unsaturated zones with an elevated value of van Genuchten parameter ($n$) making a steep water retention curve [**Fig. (9)**]. This steep curve is close to the sharp front approximation of the single front model and consequently, the single front model work well for this type of soils.

On the other hand, the Guelph loam has a water retention curve with large transition zone and the approximation of this relation by the single front model didn't give a good agreement.

As a summary, for highly dynamic boundary condition, the single front model approximation has reasonable results in the case of soils that compatible to the original Green-Ampt assumptions. However, it may not be so useful for soils that are far away from these assumptions.

### 5.3. Performance of the two-front model

In this section, we consider the performance of the two-front model. **Fig. (12) and (13)** show the efficiency of this novel two-front model for the fine sand and the Guelph loam respectively.

A qualitative comparison of the results with the previous single front model shows that the two-front approximation give a better result either in terms of the water table or the bottom flux fluctuations. The modifications/improvements are more obvious in the case of the Guelph loam compared to the previous one-front model.

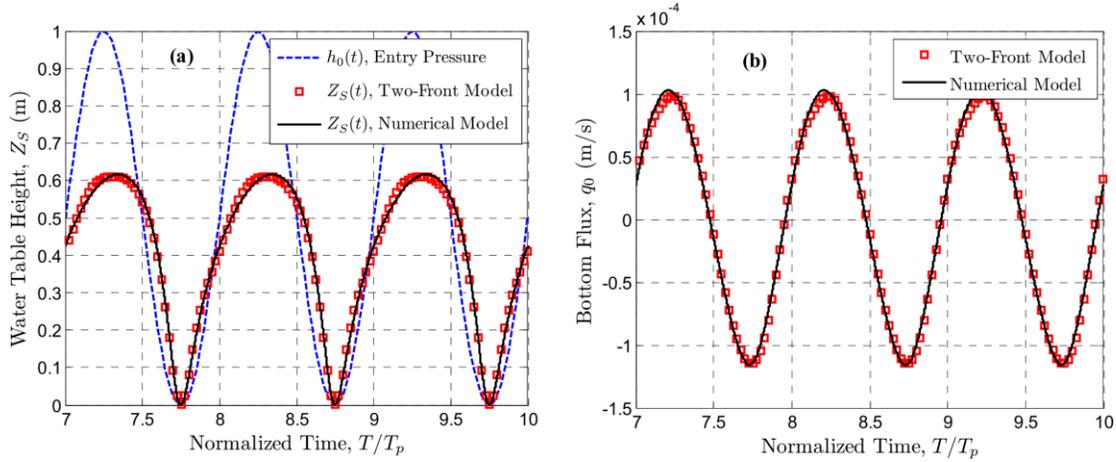

**Fig. 12:** Tidal oscillations in the SilicaSand: two-front model vs. Richards. (a) Evolution of the water table height $Z_S(t)$; (b) Evolution of the bottom flux $q_0(t)$. Bold line: Quasi-exact solution obtained from numerical solution of Richards equation; squares symbols: the two-front model. The parameters of the entry bottom pressure $h_0(t)$ are: $A_0$ =0.5m, $\overline{h_0}$ =0.5m and $T_p$ =600s. The suction head at the two fronts are calculated with eqs. 27, 28 and 29.

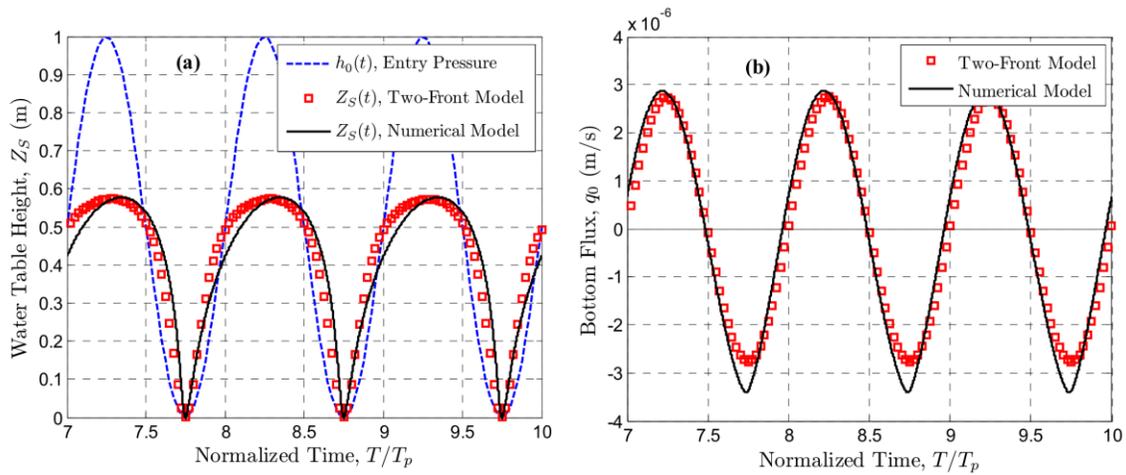

**Fig. 13:** Tidal oscillations in the Guelph Loam: two-front model vs. Richards. (a) Evolution of the water table height $Z_S(t)$; (b) Evolution of the bottom flux $q_0(t)$. Bold line: Quasi-exact solution obtained from numerical solution of Richards equation; squares symbols: the two-front model. The parameters of the entry bottom pressure $h_0(t)$ are: $A_0$ =0.5m, $\overline{h_0}$ =0.5m and $T_p$ =600s. The suction head at the two fronts are calculated with eqs. 27, 28 and 29.

Our expectations that the results can be improved more if we optimized the two-front model parameters. For this purpose, an optimization procedure was used to look for the best model parameter ($\psi_{Inter}$) to reduce the root mean square (RMS) of the difference between the two-front model and the true (numerical) model.

The two-front model was coupled with the MatLab genetic algorithm code. **Fig. (14)** shows the improved results that can be obtained with the optimized parameters for the Guelph loam.

Therefore, the successful performance of the two-front model depends largely on the appropriate values of the model parameters.

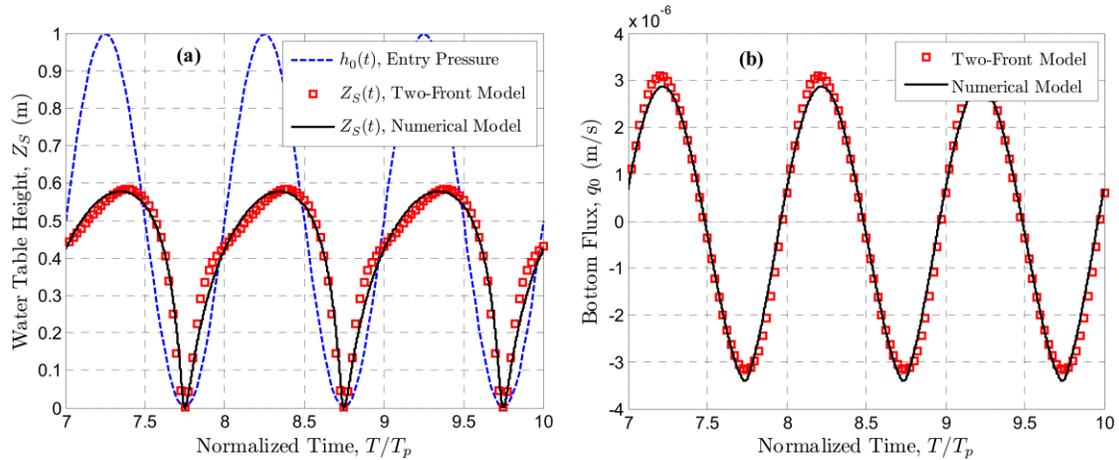

**Fig. 14:** Tidal oscillations in the Guelph Loam: two-front model with optimized parameters vs. Richards. (a) Evolution of the water table height $Z_S(t)$; (b) Evolution of the bottom flux $q_0(t)$. Bold line: Quasi-exact solution obtained from numerical solution of Richards equation; squares symbols: the optimized two-front model. The parameters of the entry bottom pressure $h_0(t)$ are: $A_0$ =0.5m, $\overline{h_0}$ =0.5m and $T_p$ =600s.

### 5.4. Performance of the Multi-front model

**Fig. (15)** and **Fig. (16)** show the evolution of the water table and the bottom fluxes as obtained from the multi-front and numerical models.

A full agreement was achieved between the multi-front model and the true solution. The multi-front model could be used as a substitute for the numerical solution of the Richards equation in the case of 1D simulation subjected to highly dynamic boundary conditions with a successful performance. Overall, our results indicate that this excellent performance can be achieved regardless of the type of saturated/unsaturated hydraulic characteristics of the porous media (coarse sand, fine sand, loam, etc.).

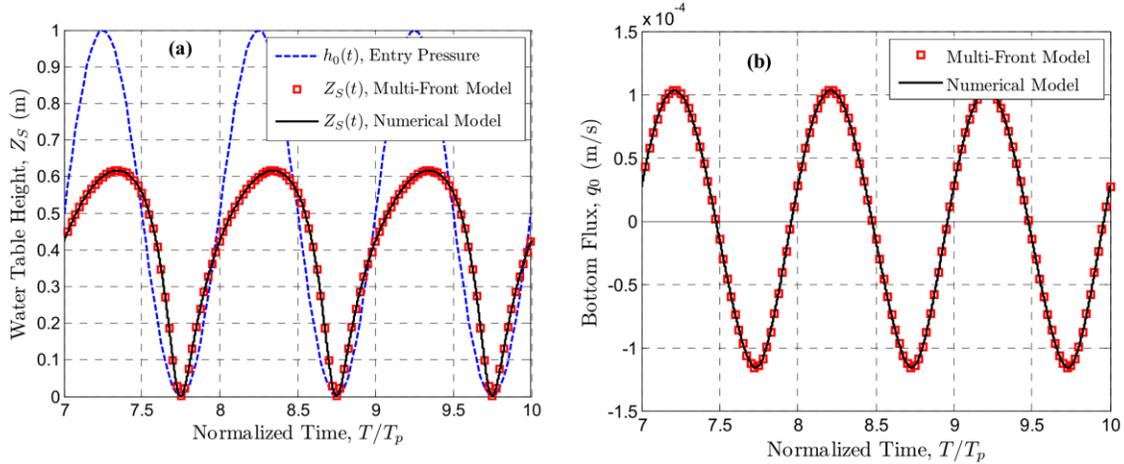

**Fig. 15:** Tidal oscillations in the SilicaSand: multi-front model vs. Richards. (a) Evolution of the water table height $Z_S(t)$; (b) Evolution of the bottom flux $q_0(t)$. Bold line: Quasi-exact solution obtained from numerical solution of Richards equation; squares symbols: the multi-front model (N=20 fronts were used). The parameters of entry bottom pressure $h_0(t)$ are: $A_0$ =0.5m, $\overline{h_0}$ =0.5m and $T_p$ =600s.

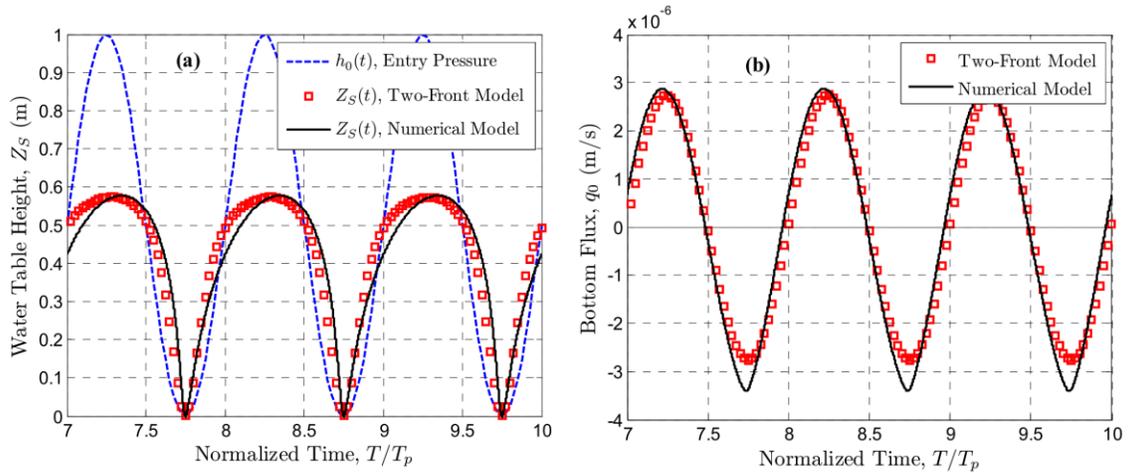

**Fig. 16:** Tidal oscillations in the Guelph Loam: multi-front model vs. Richards. (a) Evolution of the water table height $Z_S(t)$; (b) Evolution of the bottom flux $q_0(t)$. Bold line: Quasi-exact solution obtained from numerical solution of Richards equation; squares symbols: the multi-front model (N=20 fronts were used). The parameters of entry bottom pressure $h_0(t)$ are: $A_0$ =0.5m, $\overline{h_0}$ =0.5m and $T_p$ =600s.

## 6. Conclusions and outlook

In summary, a semi-analytical multi-front model which generalizes the Green-Ampt piston flow approach was developed and tested for vertical oscillatory flows in this paper. The set up is a partially saturated vertical column, including a water table, tidally forced via a sinusoïdal pressure signal imposed at the bottom of the column (experimentally, this can be realized using a tide machine as described in **Fig. 1**). In this paper, we focus on large amplitude oscillations. The amplitude of entry pressure head is up to 100% relative to the static height of the water table (most of the tests presented in this paper are based on this extreme case).

In the context of beach hydrodynamics, the objective is to analyze the non-linear time response or the frequency response of the system (the vertical soil column) in terms of pressure, moisture, water table elevation and flux (profiles and signals). Indeed, it was shown that the N-front model, with reasonably small N on the order of 10 or even less (depending on soil type) yields accurate solutions in terms of output flux signal $q_o(t)$ and water table signal $Z_s(t)$.

Furthermore, other numerical investigations indicate that the multi-front model can also give accurate solutions, not only in terms of water table and flux signals, but also in terms of vertical profiles θ(z,t) and h(z,t), and this for all types of soils. An example is shown in **Fig.17** for the case of the Guelph Loam, with N=20 fronts. In fact, the fit is almost as good for N=10 instead of 20 fronts. Many other cases (not shown here) indicate that the 20-front solution profiles are indeed quite close to the numerical profiles obtained from a finite volume solution of Richards equation, with about 200 nodes for vertical discretization in the latter. Moreover, the 20-front solution requires only 1/10th the CPU time of the 200 nodes finite volume solution, which makes it easier to conduct parametric studies with the multi-front method.

For example, **Fig. 18** shows one result from a parametric study of frequency response, conducted on the Guelph Loam using the 20-front model. The mean water table height Mean(Zs) is plotted versus the period (Tp) of the oscillating pressure $h_0(t)$ imposed at the bottom of the column.

Remarkably, in **Fig. 18**, it can be seen that the behavior of Mean (Zs) vs. Tp is non monotonic, pointing out the existence of a characteristic frequency for that soil. In other words, there appears a characteristic frequency that separates low / high frequency regimes for this soil (in **Fig. 18**, the characteristic period Tp is about 1800 s, or 1/2 hour).

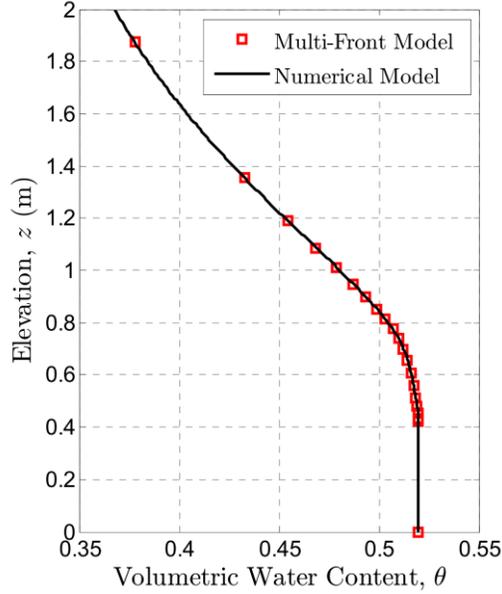

**Fig. 17:** Water content profile θ(z,t) at time t = 3000s = $5T_p$ for the Guelph Loam. The squares symbols represent the multi-front model with N=20 fronts. The solid line is the implicit finite volume solution of Richards equation obtained numerically with the Bigflow code. The parameters of the periodic bottom pressure are: $A_0$ =0.5m, $\overline{h_0}$ =0.5m and $T_p$ =600s.

Note also that the mean water table height remains below the static level $\overline{h_0}$ =0.50 m for all frequencies (or equivalently, for all periods Tp). Thus, we observe here an "under-elevation" phenomenon due to vertical capillary effects. This effect is most probably influenced by the specific type of geometry and boundary condition used in the present study (namely, vertical oscillations are forced in the system by applying an oscillating pressure at the bottom). With other geometries, however, an over-elevation of the mean water table height has been observed in the presence of lateral tidal forcing and sloping soil surface [5,39,40].

The high and low frequency limits of mean Zs are also of interest. It appears from **Fig. 18** that the infinite frequency limit for the mean water table height is the static level 0.50 m - although physically, the infinite frequency limit is not strictly valid because acceleration terms are neglected in the Darcy-Richards model used in this work. It is quite clear, on the other hand, that the mean water table height attains the static value $\overline{h_0}$ in the limit of very low frequency (Tp → ∞).

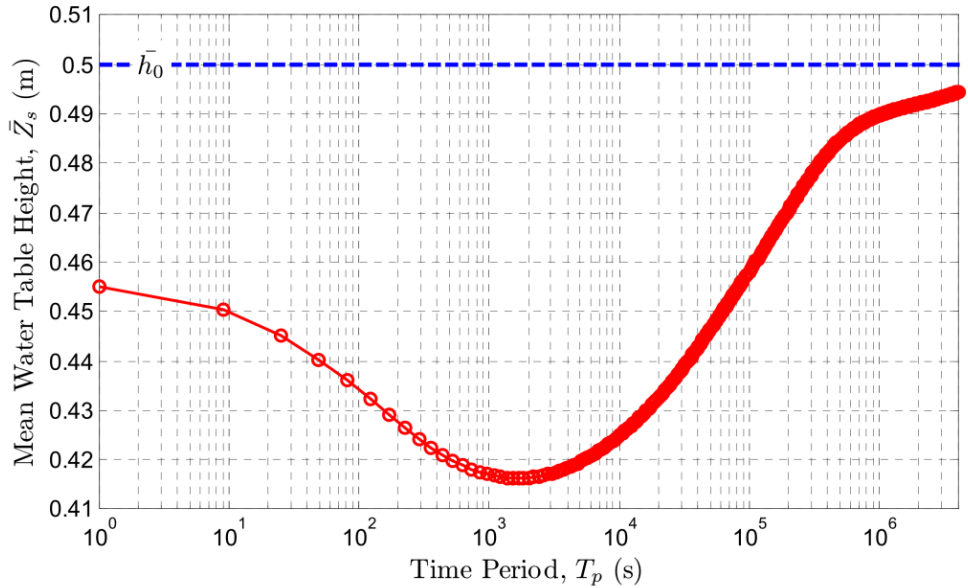

**Fig. 18:** Example of frequency response analysis using the multi-front model (here with N=20 fronts). This figure shows a plot of the mean water table height (Mean (Zs)) vs. period ($T_p$) for the Guelph Loam. The period $T_p$ (abscissa) is that of the oscillating pressure $h_0(t)$ imposed at the bottom of the column. In this case, the amplitude of $h_0(t)$ was $A_0$=0.5m, and the mean static pressure was $\overline{h_0}$ =0.5m.

The multi-front model is currently being exploited along these lines, towards a more systematic investigation and parametric study of the frequency response of soil columns in response to tidal forcing, depending on tidal amplitude, static level, and soil parameters such as Ks, θs, and capillary length scale (ongoing work).

Another interesting feature of the multi-front approach developed in this paper is the simple 2-front model. It was demonstrated that the 2-front model can give reasonable solutions - at least compared to the single front Green-Ampt model - in terms of entry flux qo(t) and water table height Zs(t). Thus, because of its simplicity, and given the favorable results of tests presented in this paper, it is thought that the 2front model is useful in its own sake as an analytical tool for studying water table oscillations in the presence of capillary effects (ongoing work).

Finally, other extensions have been considered for the generalized G-A or multi-front approach. The 2019 paper by Alastal et al. [38] applies the multi-front method to a variety of transient vertical flow problems with *non periodic forcing*, such as infiltration to a water table, sudden capillary rise, or gradual water table rise (in the latter two cases the flow is cuonter-gravitational). Other extensions to be considered could involve multidimensional geometries, including: soil surface slopes, bedrock slopes, slanted river banks, and internal heterogeneities (e.g., soil layers, cavities or galleries, beach drainage systems, etc.): however these would require serious modifications or extensions of the multi-front approach developed here and/or in Alastal et al. 2019 paper [38]. Recall that classical Green-Ampt infiltration is equivalent to a single-front of the present multi-front method. Extensions of the classical Green-Ampt approach (i.e. single front approach) have been developed in the literature and could perhaps be used for extensions of our present multi-front approach. See for instance [25,41] concerning Green-Ampt infiltration approaches for stratified soils, [28] concerning Green-Ampt infiltration with a sloping surface. See also [19] concerning a Green-Ampt correction of the planar groundwater flow equation, combining horizontal groundwater flow with vertical exchange through the unsaturated zone above the water table. Finally, in beach hydrodynamics, with tidal forcing of subsurface water, previous investigations [40] clearly show that several questions still remain open, e.g., concerning the mean over-elevation of water tables compared to static levels. Indeed, several factors intervene together: multidimensional geometry, beach slope, but also, capillary effects during unsaturated wetting/drainage cycles. A multi-front approach might be developed for studying such multidimensional oscillatory systems.

It is thought that some of the above-mentioned approaches - dealing with geometric or multidimensional extensions of the Green-Ampt model - might be useful in the context of coastal and beach hydrodynamics, and that they can be enhanced based on the two-front and multi-front methods tested in this work.